\documentclass[12pt]{article}

\usepackage{epsfig}
\usepackage{amssymb} 
\usepackage{amsmath} 
\usepackage{amsfonts} 
\usepackage{cite} 

\newcommand{\Pom}{\mathbb{P}}
\newcommand{\be}{\begin{equation}}
\newcommand{\ee}{\end{equation}}
\newcommand{\bea}{\begin{eqnarray}}
\newcommand{\eea}{\end{eqnarray}}
\newcommand{\wid}{0.8\columnwidth}

\newcommand{\kt}{\vec{k}}
\newcommand{\p}{\vec{p}}

\begin{document}

\begin{titlepage}

\begin{flushright}
\begin{tabular}{l}
 CPHT--RR005.0106  \\
 LBNL--59084 \\
\end{tabular}
\end{flushright}
\vspace{1.5cm}

\begin{center}

{\LARGE \bf
Transversity GPD in photo- and electroproduction of two vector mesons}

\vspace{1cm}

{\sc R.~Enberg}${}^{1,2}$,
{\sc B.~Pire}${}^{1}$,
{\sc L.~Szymanowski}${}^{3,4,5}$ 
\\[0.7cm]
${}^1${\it
CPhT, {\'E}cole Polytechnique, F-91128 Palaiseau, France\footnote{
  Unit{\'e} mixte 7644 du CNRS.} \\[0.1cm]
${}^2$ {\it
Lawrence Berkeley National Laboratory, Berkeley, CA 94720, USA   } \\[0.1cm]
${}^3$ {\it
 So{\l}tan Institute for Nuclear Studies,
Ho\.za 69, 00-681 Warsaw, Poland
                       } \\[0.1cm]
${}^4$ {\it
Universit\'e  de Li\`ege,  B4000  Li\`ege, Belgium  } \\[0.1cm]
${}^5$ {\it
LPT$\;$\footnote{Unit{\'e} mixte 8627 du CNRS}, 
Universit{\'e} Paris-Sud, 91405 Orsay, France
                       } \\[1.0cm]
}


\end{center}
\vskip1cm
The chiral-odd generalized parton distribution (GPD), or transversity GPD, of the nucleon can be accessed experimentally through the photo- or electroproduction of two vector mesons on a polarized nucleon target, 
$ \gamma^{(*)}  N \to \rho_1 \rho_2 N'$, where $\rho_1$ is produced at large 
transverse momentum, $\rho_2$ is transversely polarized, and the mesons are separated by a large rapidity gap.
We predict the cross section for this process for both transverse and longitudinal $\rho_{2}$ production. To this end we propose a model for the transversity GPD $H_{T}(x,\xi,t)$, and give an estimate of the relative sizes of the transverse and longitudinal $\rho_{2}$ cross sections. We show that a dedicated experiment at high energy should be able to measure the transversity content of the proton.
\vskip1cm

\vspace*{1cm}

\end{titlepage}

\section{Introduction}
\begin{figure}[t]
\centerline{\epsfxsize8.0cm\epsffile{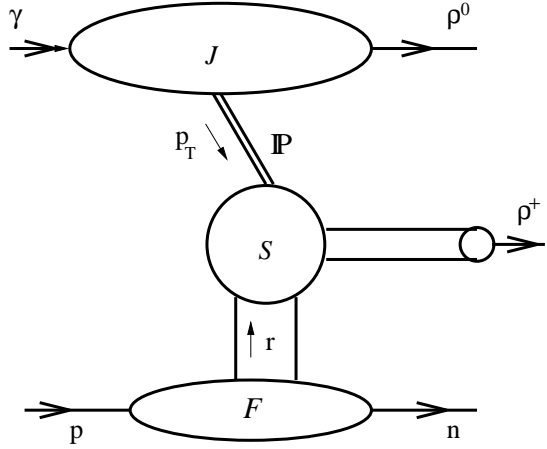}}
\caption[]{\small
Factorization of the process $\gamma^{(*)}_{L/T}\;p \to \;\rho^0_L\;\rho^+_{L/T}\;n$ in the asymmetric
kinematics discussed in the text. ${\Pom}$ describes the two gluon exchange (Born approximation of the hard pomeron).
 }
\label{fig:1}
\end{figure}

Accessing the chiral-odd quark content of the proton has a long history \cite{tra} 
which has been detailed elsewhere\cite{review}. The particular case of the chiral-odd
generalized parton distributions is interesting in itself \cite{COGPD}. Following the previous 
work in Ref.~\cite{IPST}, we consider here the process\footnote{The process $\gamma^* p \to \rho_1^0 \rho_2^0 p$ may equally well be discussed along the same lines, since its amplitude is described by exactly the same graphs, thanks to charge conjugation invariance which forbids contribution of the gluonic GPDs.}
\be
\gamma^{(*)}_{L/T} p \to \rho^0_L \rho^+_{L/T} n
\label{process}
\ee
shown in Fig.~\ref{fig:1}, that is,  virtual or real photoproduction on a
proton $p$, which leads via two-gluon exchange to the production
of  a longitudinally polarized vector meson $\rho^0_L$ separated by a large
rapidity gap from another longitudinally or transversely polarized vector meson $\rho^+_{L/T}$ 
and the scattered neutron $n$.
We consider the kinematical region where the rapidity gap between $\rho^+$ and
$n$ is much smaller than the one between
$\rho^0$ and $\rho^+$, that is the energy of the system ($\rho^+ - n$) is smaller
than the energy of the system ($\rho^0 - \rho^+$) but,  to
justify our approach, still larger
than baryonic resonance masses.

We have previously shown \cite{IPST} that in  such kinematical
circumstances the Born term for this
process is calculable consistently using
the collinear QCD factorization method. The final result is represented as an
integral (over  the longitudinal momentum fractions of the quarks)  of
the product of two amplitudes. The first one
 describes the transition $\gamma^{(*)} \to \rho^0_{L}$ via two-gluon exchange 
which can be also viewed as the Born approximation of a hard Pomeron.
The second one  describes the two gluon (Pomeron)--proton subprocess
${\Pom} p \to  \rho^+ n$ which is
closely related to the electroproduction process $\gamma^*\,N \to \rho \,N'$
where  collinear factorization
theorems allow separating  the long distance dynamics  expressed
through the
GPDs from a perturbatively calculable coefficient function. The hard scale
appearing in the process in Fig.~\ref{fig:1} is supplied by the
relatively large  momentum transfer
in the two-gluon channel, i.e. by the virtuality of the ``Pomeron''.

The first process we will calculate is the one with both vector mesons longitudinally polarized,
\be
\label{2mesongen}
\gamma^{(*)} (q)\;  p (p_2) \to  \rho_L^0(q_\rho)\;  \rho_L^+(p_\rho)\;
n(p_{2}')\;,
\ee
which involves the emission of two gluons
 in the $\gamma \to \rho_L$ transition .
We choose a charged vector meson $\rho^+$ to select quark antiquark exchange
with the nucleon line.
The second process we are interested in is the one involving the chiral-odd GPD,
 e.g. 
\be
\label{2mesontr}
\gamma^{(*)}(q)\;  p (p_2) \to  \rho_L^0(q_\rho)\;  \rho_T^+(p_\rho)\;
n(p_{2}')\;,
\ee
which is the main motivation for the study of this two meson production process.

\section{Kinematics}

%
%

Let us first summarize the details of the kinematics of the process, 
restricting to real photoproduction.
We introduce two light-like Sudakov vectors
$p_{1}$ and $P=1/2(p_2 +p_{2'})$. We also introduce the auxiliary variable
${\cal S}=2p_1\,P$, related to the total  center-of-mass energy squared of the 
$\gamma^*p-$system, $s=(q+p_2)^2$, as $s+Q^2=(1+\xi){\cal S}$.  
The momenta are parametrized as follows~:
\bea
&& q^\mu = p_1^\mu  - \frac{Q^2}{{\cal S}}P^\mu \;, 
\nonumber \\
&& q_\rho^\mu = \alpha p_1^\mu +
\frac{\p^{\;2}}{\alpha {\cal S}}P^\mu + p_T^\mu \;,\;\;\;\;\;p_T^2 =
-\p^{\;2}\;,
\nonumber \\
&& p_\rho^\mu = \bar \alpha p_1^\mu 
+ \frac{\p^{\;2}}{\bar \alpha {\cal S}}P^\mu
- p_T^\mu \;,\;\;\;\;\;\bar \alpha \equiv 1-\alpha\;, \nonumber \\
&& p_2^\mu = (1+\xi)P^\mu\,,\;\;\;\;\;p_{2\;'}^\mu = (1-\xi)P^\mu\;, 
\label{Sud}
\eea
where $Q^2=-q^2$ is the photon virtuality, and 
$\xi$ is the skewedness parameter  which can be written in
terms of the invariant mass $s_1$ of the two mesons as 
\be
\xi = \frac{s_1+Q^2}{2{\cal S}}, ~~~~~~~~~~~
s_1 = (q_\rho + p_\rho)^2 = \frac{\p^{\;2}}{\alpha \bar \alpha}.
\label{s1}
\ee
The $\rho^+$-meson--target invariant mass equals
\be
s_2 = (p_\rho + p_{2'})^2 = {\cal S}\, \bar \alpha\,\left(1-\xi  \right)\,.
\label{s2}
\ee
The kinematical limit with a large rapidity gap between the two mesons in the
final state is
obtained by demanding that $s_1$ be very large,  of the order of 
${\cal S} \approx s$,
\be
s_1 =2{\cal S}\, \xi\,,\;\;\;\;s_1 \gg \p^{\;2}\,,
\label{s1gap}
\ee
whereas $s_2$ is kept of the order of $\p^{\;2}$,
\be
\label{s2gap}
s_2 \to \frac{\p^{\;2}}{2\xi}\,\left(1-\xi  \right) \,,
\ee
large enough to justify the use
of  perturbation theory in the
collinear subprocess ${\Pom} p \to \rho^+ n$ and the application of
the GPD framework.

In terms of the longitudinal fraction $\alpha$ the limit
with a large rapidity gap corresponds
to taking the limits
\be
\label{alphagap}
\alpha \to 1\,,\;\;\;\;\;\bar \alpha s_1 \to \p^{\;2}\,,\;\;\;\;\;\;\xi
\sim
1\,.
\ee
We choose the kinematics so that the nucleon gets no transverse
momentum in the process. The $t$-dependence of the 
GPDs is not known in detail, but the sum rules which relate it to the nucleon form
factors imply a strong decrease of the cross section
with increasing $-t$. Thus, taking $t=t_{min}$ yields simpler formulas and amounts to 
most of the cross section.
One may however allow a finite
momentum transfer, small with respect to $|\p|$, correcting 
expressions of the amplitude
with contributions containing additional GPDs.

Let us repeat that the role of the main hard scale in the
processes under discussion below is played by the virtuality $p^2=-\p^2$,
which is the large momentum transfer
in  the two-gluon exchange channel. 

The 3-particle phase space needed to obtain the differential cross section is computed through the standard recurrence relation, which gives
\be
\text{dPS}_{3}(s, p_2, p_\rho, p_{2'}) = \frac{1}{256 \pi^4 s s_2} dp_{T}^2 \, dt \, ds_{2} d\Phi\;,
\ee
where $\Phi $ is the angle between the ($\gamma \rho^0$) plane and the (p, n, $\rho^+$) 
plane. At $t=t_{min}$, this angle is irrelevant and the phase space factor simplifies to
\be
\text{dPS}_{3}(s, p_2, p_\rho, p_{2'}) 
= \frac{1}{128 \pi^3 \, s\, \xi(1-\xi)}  dp_{T}^2 \, dt \, d\xi \;.
\ee
The differential cross section at the minimal value of $-t=-t_{min}$ then reads
\be
\label{crosssec}
\frac{d\sigma}{dp_T^2\,dt\,d\xi} = \frac {1} {256 \pi^3\, \xi(1-\xi) s^2} \left|{\cal M}\right|^2.
\ee

\section{The scattering amplitude}
The scattering amplitude ${\cal M}$ of the
process
(\ref{2mesongen})  using the standard collinear QCD factorization
method is written  in a form suggested by Fig.~\ref{fig:1} as:
\be
\label{fact}
{\cal M} \sim \sum\limits_{q=u,d}\int\limits_0^1 dz\,\int\limits_0^1
du\,\int\limits_{-1}^1 dx\,
T^q_H(x_1,u,z)\,H^q(x,\xi,0) \phi_{\rho^+}(u)
\phi_{\rho^0}(z)\;.
\ee
Here $H^q(x,\xi,0)$ is the generalized parton distribution of parton $q$
in the target at zero momentum transfer and $x+\xi$ and $\xi -x$ are (see
Fig.~\ref{fig:1}) the momentum
fractions of the quark and antiquark emitted by the target
(since it turns out that our kinematics selects $x < \xi$, the second parton is 
interpreted as an emitted antiquark). 
$\phi_{\rho^+}(u)$ and $\phi_{\rho^0}(z)$ are the distribution
amplitudes (DA) of the $\rho^+$-meson and
$\rho^0$-meson, respectively, and $u$ and $z$ are the corresponding lightcone momentum fractions of the quark in the meson. We will also use the shorthand 
notation $\bar z=1-z$ and $\bar u=1-u$ in the following.
$T^q_H(x,u,z)$ is
the hard scattering amplitude (the coefficient function).
For clarity of notation we omit in Eq.~(\ref{fact}) the factorization scale
dependence of $T^q_H$, $H^q$, $\phi_{\rho^0}$ and $\phi_{\rho^+}$.

Eq.~(\ref{fact}) describes the amplitude in the leading twist
approximation. In other words
all terms suppressed by powers of a hard scale parameter $1/|\p|$ are
omitted.  Within this approximation one neglects (in the physical gauge)
the contributions of higher Fock states in the meson wave functions
and many parton correlations (higher twist GPDs) in the proton.
Moreover, we also use the collinear approximation:\ in the hard scattering amplitude we neglect the relative
transverse momenta (with respect to the meson momentum) of the constituent
quarks. These approximations result in the appearance of the distribution amplitudes in the factorization formula (\ref{fact}),
i.e. the lightcone wave functions, integrated over 
the relative transverse
momenta of constituents up to the collinear
factorization scale.

An additional simplification appears in the kinematics given by
Eqs.~(\ref{s1gap}--\ref{alphagap}). In this limit one
needs consider only the diagrams which involve two-gluon exchange between the two
mesons. The other contributions (the fermion exchange diagrams) to the
coefficient function $T_H^p$
are known \cite{BFKL} to be suppressed by powers of $
\p^{\;2}/s$. Therefore we
will
not discuss them here. At the same accuracy,
i.e.\ neglecting
terms $\sim \p^{\;2}/s$, the contribution of gluon exchange diagrams shown
in Fig.~2 
turns out to be
 purely
imaginary. It involves GPDs in the ERBL region $-\xi \, <\,x\, <\, \xi$ only, which is quite specific for our process \cite{IPST}.

The amplitude may then be written in terms of the impact factor
$J^{\gamma^{(*)}_{L/T} \to \rho^0_L}$ as
\bea
\label{CEN}
&&{\cal M}^{\gamma^{(*)}_{L/T}\,p\,\to \rho_L^0\, \rho^+_L\,n}=
i 16\pi^2   s \alpha_s f_\rho^+ \xi \sqrt\frac{1-\xi}{1+\xi}
\frac{C_F}{N\,(\p^{\;2})^2}
\nonumber \\
&&\times\int\limits_0^1
\frac{\;du\;\phi_\parallel(u)}{ \,u^2 \bar u^2 }
 J^{\gamma^{(*)}_{L/T} \to \rho^0_L}(u\p,\bar u\p)
 \left[ H^u(\xi (2u-1),\xi,0)- H^{d}(\xi (2u-1),\xi,0)\right].
\eea
The impact
factor  $J^{\gamma^{*}_L \to \rho^0_L}$ has the form
\be
\label{ifgamma}
J^{\gamma^{(*)}_L \to \rho^0_L}(\kt_1,\kt_2)=  -   f_\rho \frac{e
\alpha_s 2\pi
  Q}{N_c\sqrt{2}} \int\limits_0^1 dz\;z\bar z 
\phi_{||}(z)P(\kt_1,\kt_2)\;,
\ee
with the $\phi_{||}(z)$ DA defined by the matrix element
\be
\label{phi||}
\langle 0 | \bar q(0) \gamma^\mu q(y)|\rho^0_L(q_\rho)\rangle 
= q_\rho^\mu f_{\rho}^0 \int\limits_0^1dz\;e^{-iz(q_\rho y)}\phi_{||}(z)\;. 
\ee 
For $\gamma^{(*)}$ transversely polarized, 
$J^{\gamma^{(*)}_T \to \rho^0_L}$ reads:
\be
\label{TL}
J^{\gamma^{(*)}_T \to \rho^0_L}(\kt_1,\kt_2=\p-\kt_1)=
-\frac{e\,\alpha_s\,\pi\,f_\rho^0}{\sqrt{2}\,N}\;\int\limits_0^1\,dz\,(2z-1)\,
\phi_\parallel(z)\,\left( \vec{\varepsilon}\,\vec Q_P \right)\;.
\ee
Here $\vec{\varepsilon}$ is the polarization vector of the initial photon,
\bea
\label{P}
P(\kt_1,\kt_2=\p-\kt_1)=&&\frac{1}{z^2\p^{\;2}+m_q^2 +Q^2z\bar z} +
\frac{1}{{\bar z}^2\p^{\;2}+m_q^2 +Q^2z\bar z} \nonumber \\
&&-\frac{1}{(\kt_1-z\p\,)^2+m_q^2 +Q^2z\bar z} -
\frac{1}{(\kt_1-\bar z \p\,)^2+m_q^2 +Q^2z\bar z}\;,
\eea
and
\bea
\label{Q}
&&\vec Q_P(\kt_1,\kt_2=\p-\kt_1)=
\frac{z\,\p}{z^2\,\p^2+Q^2\,z\,\bar z +m_q^2}
- \frac{\bar z\,\p}{\bar z^2\,\p^2+Q^2\,z\,\bar z +m_q^2}
 \\
&&\hspace{3cm}+\frac{\kt_1 - z\,\p}{(\kt_1 - z\,\p)^2+Q^2\,z\,\bar z +m_q^2}
-\frac{\kt_1 - \bar z\,\p}{(\kt_1 - \bar z\,\p)^2+Q^2\,z\,\bar z +m_q^2}\;.
\nonumber
\eea
In all our estimates in this paper we put the quark mass $m_q=0$.

In the case of the transversely polarized $\rho^+_{T}$ meson production, 
one instead gets
\bea
\label{CON}
&&{\cal M}^{\gamma^{(*)}_{L/T} \,p\,\to \rho_L^0\, \rho^+_T\,n}=
 -\,\sin \theta \;16\pi^2 s \alpha_s f_\rho^T \xi \sqrt\frac{1-\xi}{1+\xi}
\frac{C_F}{N\,(\p^{\;2})^2}
\nonumber \\
&&\times\int\limits_0^1
\frac{\;du\;\phi_\perp(u)}{ \,u^2 \bar u^2 }
 J^{\gamma^{(*)}_{L/T} \to \rho^0_L}(u\p,\bar u\p)
 \left[ H_{T}^u(\xi (2u-1),\xi,0)- H_T^{d}(\xi (2u-1),\xi,0)\right]\;,
\eea
 which involves the chiral-odd
transversity distribution whose investigation is the main motivation of our
studies. $J^{\gamma^{(*)}_{L/T} \to \rho^0_L}$ are the same impact factors 
as in
(\ref{CEN}) and   $\theta$ is the angle between the transverse 
polarization vector of
the target $\vec{n}$ and the polarization vector
$\vec{\epsilon}_T$ of the produced $\rho^+_T$-meson. In our numerical studies we take $\theta=\pi/2$, but this dependence should of course be confirmed experimentally.

The chiral-odd light-cone distribution amplitude for the
transversely polarized $\rho$-meson is defined
by the matrix element \cite{BalBr}
\be
\label{rho+T}
\langle \rho_T(p_\rho,T) \mid \bar q(x) \sigma^{\mu \nu} q(-x)\mid 0
\rangle =i f_\rho^T \left(p_\rho^{\mu}\epsilon^{*\nu}_T -
p_\rho^{\nu}\epsilon^{*\mu}_T
\right)
\int\limits_0^1 du e^{-i(2u-1)(p_\rho x)}\;\phi_\perp(u)\;.
\ee 
We use the asymptotic forms of both $\rho_T^+$ and $\rho^0_L$ DAs, i.e.,
$\phi_\perp(u)=\phi_\|(u)=6u\bar u$.

We use the values of the meson decay constants in the above expressions at the
scale 1~GeV:\ 
$f_\rho^+=198\pm 7\,$MeV for the longitudinal $\rho^+$, 
$f_\rho^0=216\pm 5$\,MeV for the longitudinal $\rho^0$, and 
$f_\rho^T=160\pm 10\,$MeV for the transverse $\rho^+$.

The generalized transversity distribution in the nucleon target described by the
polarization vector $n^\mu$  is
defined by the formula \cite{Diehl:2001pm}
\be
\label{FT}
\hspace*{-0.5cm}\int \frac {dz^-}{4\pi}
e^{i x P^+ z^-}
\langle N(p_{2'},n)|\bar q(-\frac{z}{2})\, i \sigma^{+\,i}\, q(\frac{z}{2})| N(p_2,n) \rangle
=\frac{1}{2P^+}\bar u(p_{2'},n) i \sigma^{+ \,i} u(p_2,n) H_{T}^q(x,\xi,t),
\ee
where terms vanishing at $t=t_{min}$ (i.e.\ $\Delta_T=0$) have been neglected. 
Our convention is $\sigma^{\mu\,\nu}=i/2[\gamma^\mu\,,\gamma^\nu]$.

Finally, two remarks are in order.
Note that as $|{\cal M}|^2$, both from Eq.~(\ref{CEN}) and Eq.~(\ref{CON}), is proportional to $s^2$, 
the $s$-dependence cancels in the full cross 
section (\ref{crosssec}). 
We want to stress that this is valid only for the high energy limit
discussed in Sec.~2. If one is willing to 
study these processes 
 at lower energy, 
one should 
include all polarization states of exchanged 
gluons\footnote{At high energies the longitudinal polarization states
of $t$-channel gluons (or in the Regge theory language the ``nonsense'' polarizations) 
give the dominant contribution and we omit the contribution of other polarization states.
} and the additional contributions related to quark exchanges. 

%
\begin{figure}[tb]
\centerline{\epsfig{file=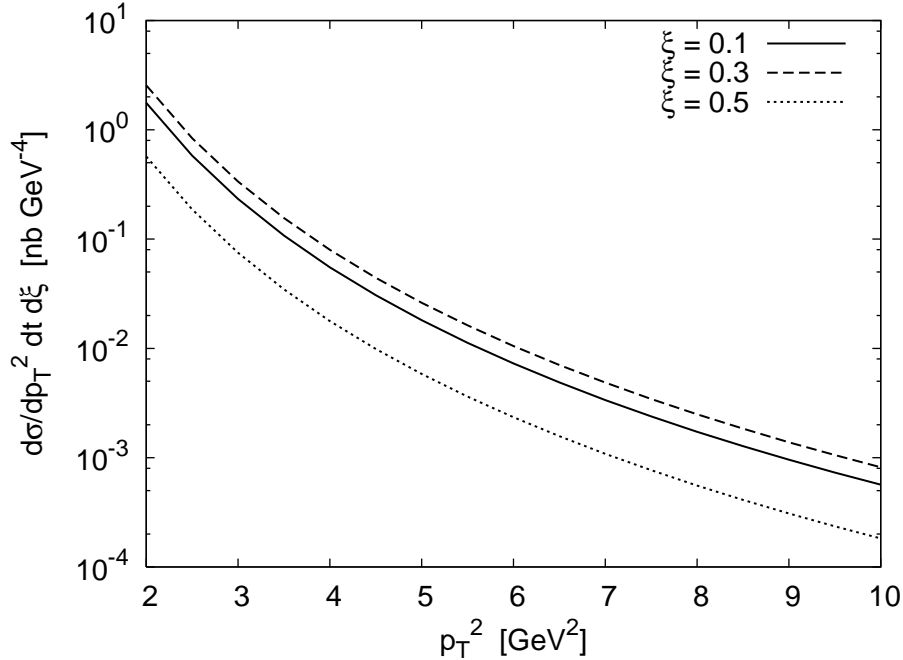,width=\wid}}
\caption{\small
The differential cross section for the photoproduction of 
longitudinally  polarized $\rho^0$ and $\rho^+$ as a function of the squared transverse momentum $p_T^2$ for $\xi=$ 0.1, 0.3, and 0.5.
 }
\label{photolong}
\end{figure}
%

%
\begin{figure}[tbh]
\centerline{\epsfig{file=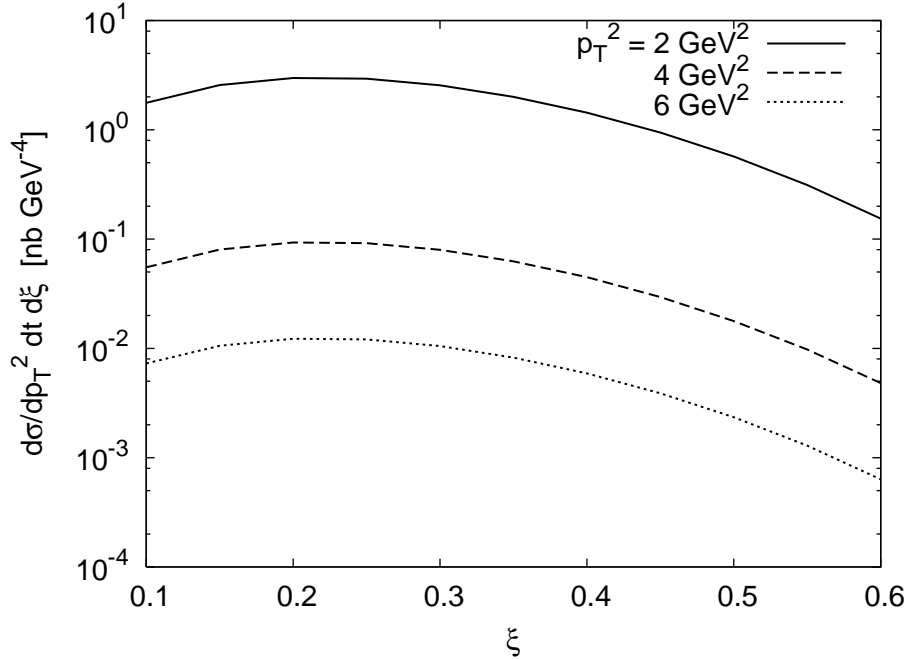,width=\wid}}
\caption{\small
The differential cross section for the photoproduction of 
longitudinally  polarized $\rho^0$ and $\rho^+$ as a function of $\xi$ for $p_T^2=$ 2, 4, and 6 GeV$^2$.
 }
\label{photolongxi}
\end{figure}
%

\section{Longitudinally polarized meson case}
 Let us first estimate the rate in the longitudinally polarized meson case. 
As in other $\rho$ meson production processes, the dominant contribution comes from the unpolarized 
GPD  $H(x,\xi, t)$ and we will neglect the $E(x,\xi,t)$ contributions which vanish at $t=t_{min}$. 
Our process selects the 
isovector part of this GPD.  
We use a standard description of this GPD in terms of double distributions, due to 
Radyushkin \cite{Radyushkin}. In general, it should be supplemented by a D-term \cite{Dterm} contribution. 
Since the chiral quark model estimates show that this latter term is almost flavor independent \cite{GPV}
we will not include it in our model calculation involving isovector GPDs. We assume thus,
\begin{align}
H(x,\xi, t) &=
\frac{\theta(\xi+x)}{1+\xi} 
\int_0^{\min\left[\frac{\xi+x}{2\xi},\frac{1-x}{1-\xi}\right]}
d y \, F^q\left( \frac{\xi+x-2\xi y}{1+\xi}, y , t \right)
\nonumber \\ 
&-\frac{\theta(\xi-x)}{1+\xi} 
\int_0^{\min\left[\frac{\xi-x}{2\xi},\frac{1+x}{1-\xi}\right]}
d y \, F^q\left( \frac{\xi-x-2\xi y}{1+\xi}, y , t \right)\;,
\end{align}
where the double distribution  $F^q(X,Y,t)$ is given by the ansatz \cite{Radyushkin}
\be
F^q(X,Y,t) = \frac{F_1^q(t)}{F_1^q(0)} q(X) 6 \frac{Y(1-X-Y)}{(1-X)^3}.
\label{doubledistro}
\ee
In the expression (\ref{doubledistro}),  $q(X)$ is the quark parton distribution function, for which we use the parametrization of Ref.~\cite{MRST}. The $t$-dependence of the GPDs is given by the functions $F_1^q(t)$, which are related to the electromagnetic form factors of the proton and neutron, but since we are only evaluating the amplitudes at $t=t_{min}$ their form is not important.

To compute the cross sections we must in general evaluate the 
integrals over $u$ and $z$ numerically; in the photoproduction case 
care must be taken to properly treat the apparent divergences at $u=z$ 
and $u+z=1$.

\begin{figure}[htb]
\centerline{%
\epsfig{file=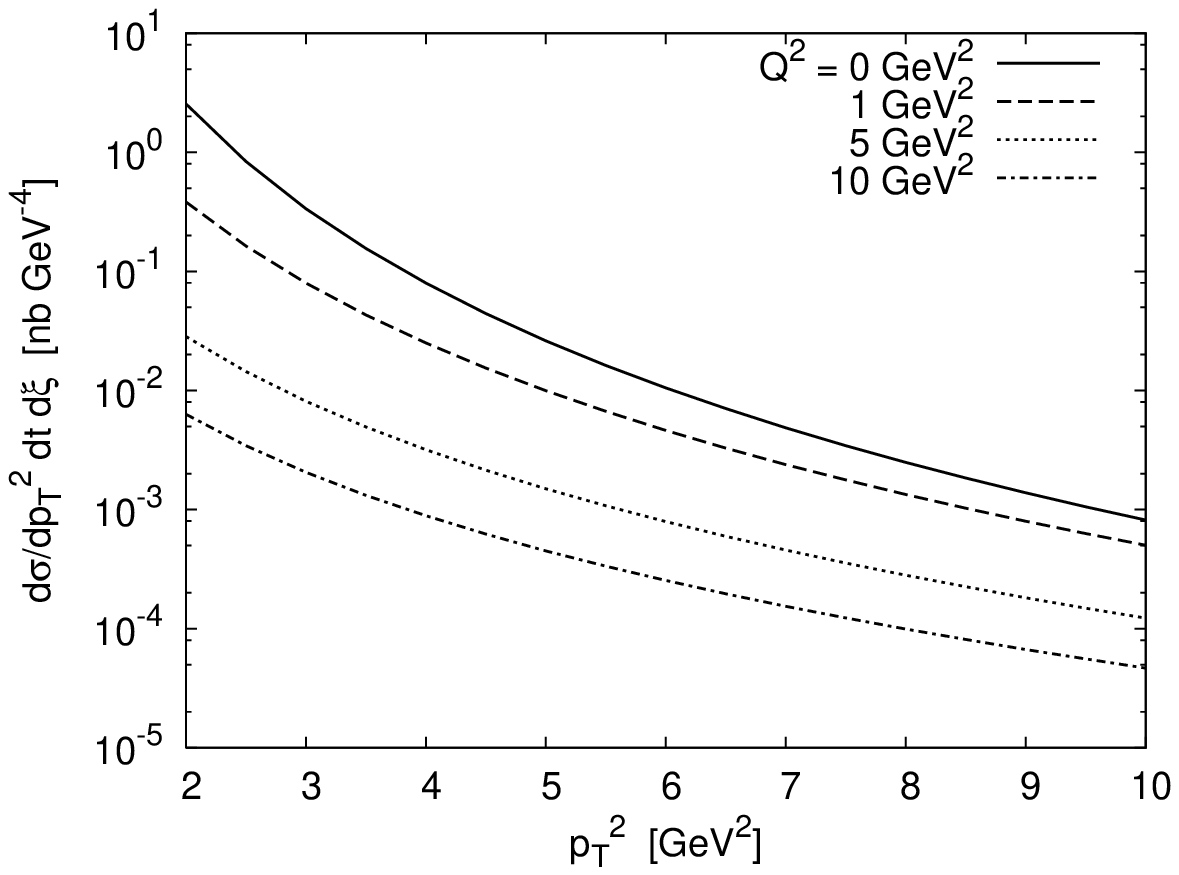,width=0.5\columnwidth}%
\epsfig{file=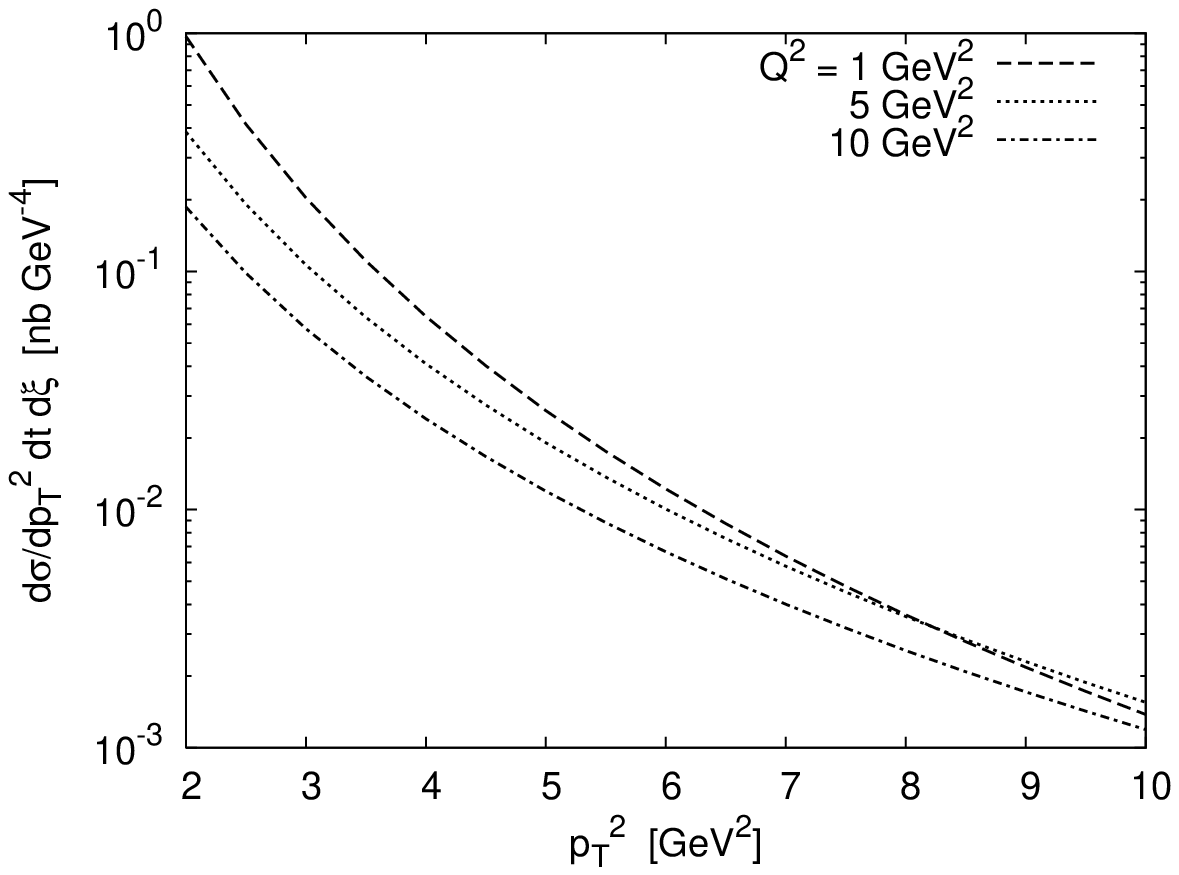,width=0.5\columnwidth}}
\caption{\small
The differential cross section for $\gamma^*_{L/T}(Q) \,p\to \rho^0_L\,\rho^+_L\,n$ for transverse virtual photon (left) and longitudinal virtual photon (right), plotted as a function of $p_T^2$ for $\xi=0.3$ and $Q^2=$ 0, 1, 5, and 10 GeV$^2$.
 }
\label{electrolong}
\end{figure}

The differential cross section for longitudinally  polarized $\rho^+$ 
production is shown in Figs.~\ref{photolong}, \ref{photolongxi} and \ref{electrolong} for the real and virtual photon cases.
We conclude from them that the photoproduction rate is much larger than the 
electroproduction rate (note that we did not include the 
additional suppression factor coming from the virtual photon flux and instead show the transverse and longitudinal photon cases separately).

%
%

\section{Modeling the transversity GPD}
To estimate the rate of our process in the case of transversely polarized
$\rho^+$ we need to formulate a model for the
transversity dependent GPD  $H_{T}(x,\xi, t)$. From Eq.~(\ref{CON}) 
it follows that we only need  a model for the transversity GPD in 
the ERBL region, 
$-\xi \leq x \leq \xi$. 
Unfortunately, even in the case of forward $h_1$ structure function only very
rough bag model estimates are available \cite{JJ}. 
The bag model estimate of the leading transversity GPD was given recently 
in Ref.~\cite{Scop}. 
Recent progress was done also with lattice methods which have estimated 
its first $x$ moments   \cite{Lattice}.

We propose a simple model inspired by Ref.~\cite{MPR} in which 
the $\tilde E$ GPD was evaluated within 
the chiral approach, where a meson exchange dominates the physical process
encoded in the GPD. 
The analogous meson pole approach
 was applied in the case of the forward transversity distribution in
Ref.~\cite{GG}. Below we generalize 
the pole model of the nucleon tensor charge developed in \cite{GG}
 to the non-forward kinematics.
 
%
\begin{figure}[htb]
\centerline{\epsfig{file=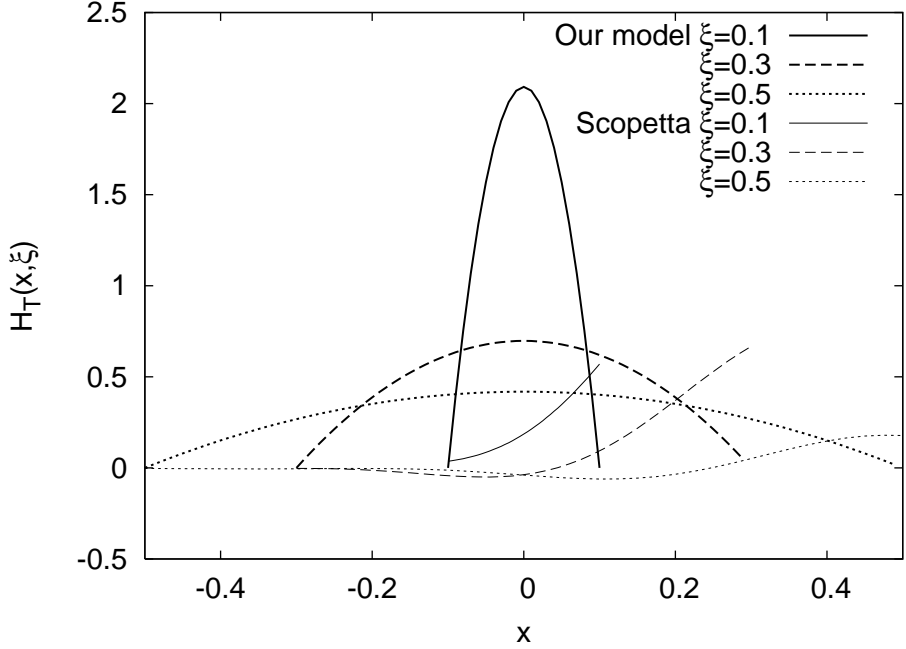,width=\wid}}
\caption{\small 
The transversity GPD $H_T(x,\xi,0)$ for various values of $\xi$ for $x$ in the ERBL region, in our model
(thicker curves) and in the model of Ref.~\cite{Scop} (thinner curves).
 }
\label{HT}
\end{figure}
%

We start with the effective interaction Lagrangian 
\be
{\cal L_{ANN}}= 
 \frac  {g_{A\, NN}}{2M}\bar N \sigma_{\mu\nu}\gamma_5
\partial^\nu A^\mu N \;,
\label{ANN}
\ee
in which $g_{A NN}$ is the coupling constant determining  the strength of 
the interaction of 
the axial meson $A$ with the nucleon $N$. Inserting the interaction term
$i\int d^4x \; {\cal L_{ANN}}$
into the S-matrix element in the left hand side of (\ref{FT}) and using
the reduction formula
one can separate the contribution to $H_T$ of the axial meson pole. 
In this way, for transversely polarized nucleons described by 
the polarization vector $S_T=(0,\vec n_\perp,0)$, using the definition (\ref{FT}) of $H^q_T$, 
we get   
the non-forward version
of Eq.~(7) of \cite{GG} as
\begin{equation}
\label{pole}
H^a_T(x,\xi)=\frac{g_{ANN}f_A^{a\perp} \left(\Delta\cdot S_T  \right)^2 }{2M_N\,m_A^2}\,
\frac{\phi_\perp(\frac{x+\xi}{2\xi})}{2\xi}\;,
\end{equation}
where $\Delta$ is the transverse part of the momentum transfer vector $r$ (see Fig.~\ref{fig:1}) and $f^{a\perp}_A$ 
is related to the $A$ meson decay constant. We have here used  
 the definition \cite{BalBr}
of the transversely polarized axial vector meson 
 distribution amplitude,
 \bea
 &&\langle 0| \bar q(-1/2\,z)\sigma^{\alpha\beta}\gamma_5 
 \frac{\lambda^a}{2}q(1/2\,z)|A(k,
 \lambda)\rangle =
 \nonumber
 \\
 \label{ADA}
 && i\,f^{a\;T}_A\left[\epsilon^\alpha(\lambda)k^\beta 
 - \epsilon^\beta(\lambda)k^\alpha  \right]
 \int\limits_0^1\;du\,e^{i/2(1-2u)k\cdot z}\;\phi^A_\perp(u)\;, 
 \eea
 in which $\epsilon(\lambda)$ is the polarization vector of meson $A$ 
 with the momentum $k$, and $\phi^A_\perp(z)$ is the distribution amplitude. 
In arriving at (\ref{pole}) it is useful to note 
that the definition (\ref{FT}) can be rewritten using the Dirac 
structure $\sigma^{+j}\gamma_5$ instead of $i\sigma^{+i}$ \cite{Diehl:2001pm}.

According to the model of Ref.~\cite{GG} we now identify the scalar product
$\left(\Delta\cdot S_T  \right)^2$ with the average of the 
intrinsic transverse 
momentum of the quarks:\ $\left(\Delta\cdot S_T  \right)^2 
\to 1/2 \langle k_\perp^2\rangle$. Also, the valence quantum number of 
$t$-channel isovector exchange 
leads to the identification of the axial meson as
$A=b_1(1235)$. In this way, using the SU(2) relation
\be
\langle n | \bar d O u | p \rangle =
\langle p | \bar u O u | p \rangle -
\langle p | \bar d O d | p \rangle ,
\ee
we obtain as the final expression for the valence 
part $H^v_T = H^u_T - H^d_T $ 
(compare with Eq.~(12) of \cite{GG})
\begin{equation}
\label{Hpole}
H^v_T(x,\xi,t)=\frac{g_{b_1 NN}f_{b_1}^{T}\langle k_\perp^2\rangle }{
2\sqrt{2}M_N\,m_{b_1}^2}\,
\frac{\phi^{b_1}_\perp(\frac{x+\xi}{2\xi})}{2\xi}\;,
\end{equation}
where, according to Eqs.~(8, 9, 10) of \cite{GG},
$$
\label{fb}
f_{b_1}^T = \frac{\sqrt{2}}{m_{b_1}}f_{a_1}\;,\;\;\;\;\;\;
f_{a_1}=(0.19\pm0.03)\mbox{GeV$^2$}\;,\nonumber
$$
\begin{equation}
g_{b_1NN}=\frac{5}{3\sqrt{2}}g_{a_1NN}\;,\;\;\;\;\;g_{a_1NN}=7.49 \pm 1.0 \;,
\end{equation}
and
\begin{equation}
\label{k2}
\langle k_\perp^2 \rangle = (0.58\; \div\; 1.0)\mbox{GeV$^2$}\;.
\end{equation}
(We use the lower value $\langle k_\perp^2 \rangle=0.58$ in our numerical estimates).
In the following we assume that the distribution amplitude
$\phi^{b_1}_\perp$ takes its  asymptotic form, $\phi^{b_1}_\perp(u)=6u\bar u$.
We show in Fig.~\ref{HT} the resulting transversity GPD $H_T(x,\xi,0)$ for various values of $\xi$.

The second model which we use in our estimates is the bag model estimate of Ref.~\cite{Scop}. In the 
 Fig.~\ref{HT} we show also the comparison of this transversity GPD with the one defined by 
Eq.~(\ref{Hpole}). We see that these models lead to quite different results which consequently 
can serve as an estimate of theoretical uncertainties of our rate estimates. We want to stress that 
quark models, such as in  Ref.~\cite{Scop}, naturally underestimate the GPDs in the ERBL region, 
since they do not include the physics of meson exchange which is at the core of our model. 

\section{Cross sections for $\rho^0_L \;\rho^+_T$ production  }

%
\begin{figure}[tb]
\centerline{\epsfig{file=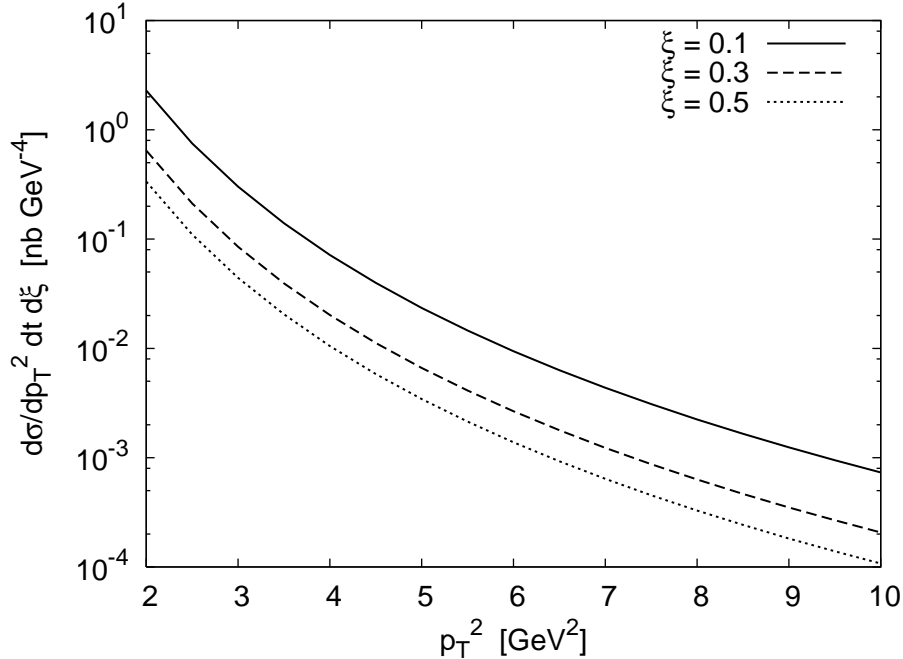,width=\wid}}
\caption{\small
The differential cross section for photoproduction of $\rho^0_L$,
$\rho^+_T$ as a function of $p_T^2$ for $\xi=$ 0.1, 0.3, and 0.5.
}
\label{Tphoto}
\end{figure}
%

%
\begin{figure}[htb]
\centerline{\epsfig{file=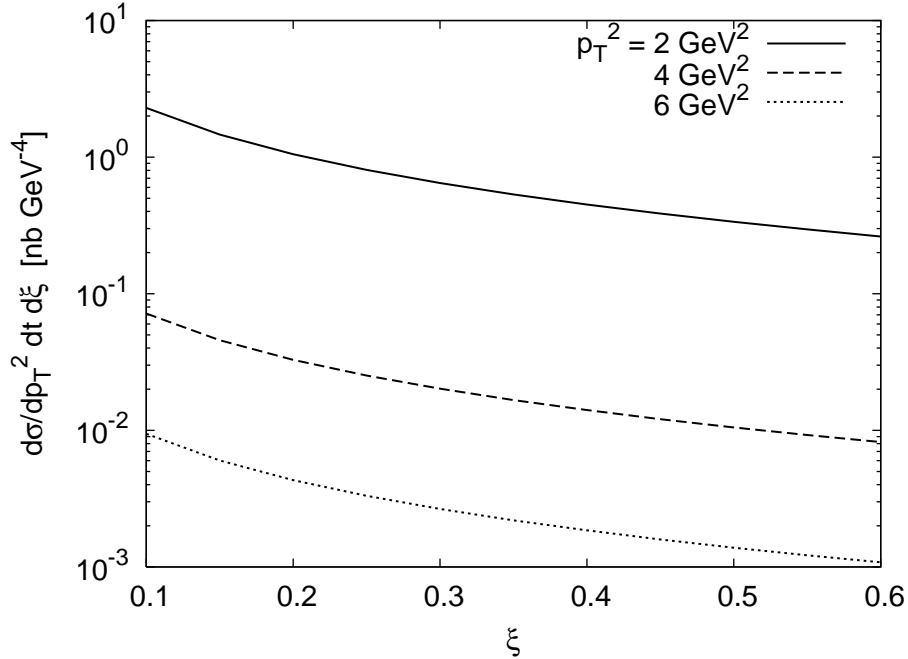,width=\wid}}
\caption{\small
The differential cross section for the photoproduction of 
transversely  polarized $\rho^0$ and $\rho^+$ as a function of $\xi$ for $p_T^2=$ 2, 4, and 6 GeV$^2$.
 }
\label{Tphotoxi}
\end{figure}
%

%
\begin{figure}[htb]
\centerline{%
\epsfig{file=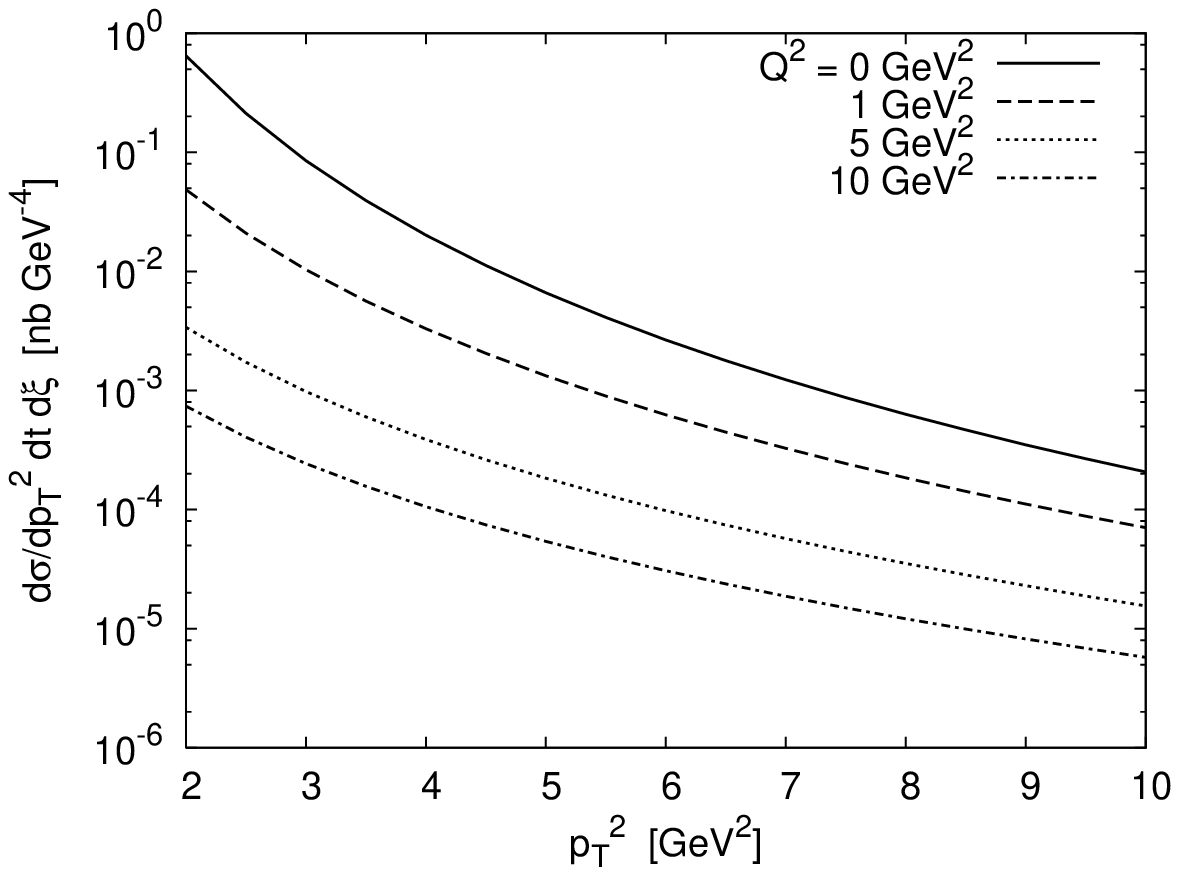,width=0.5\columnwidth}%
\epsfig{file=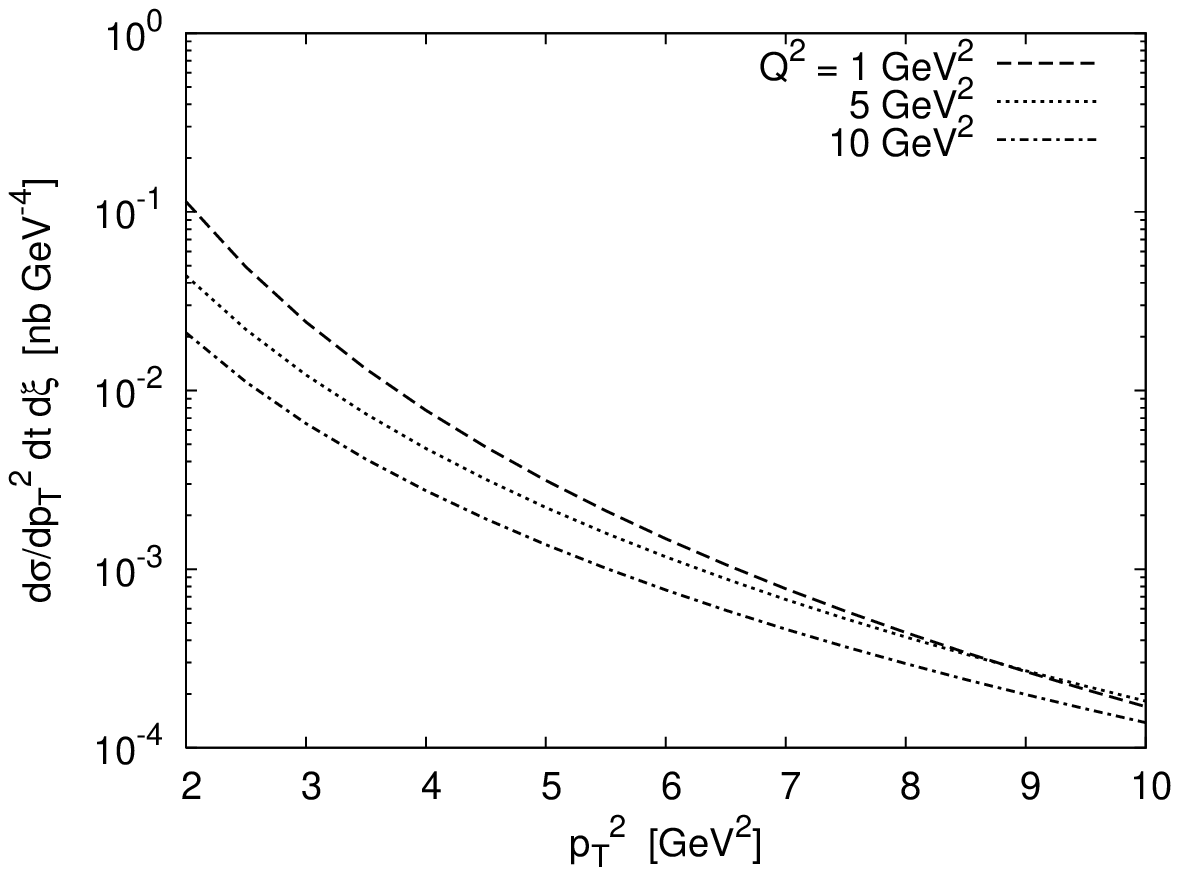,width=0.5\columnwidth}}
\caption{\small 
The differential cross section for $\gamma^*_{L/T}(Q) \,p\to \rho^0_L\,\rho^+_T\,n$ for transverse virtual photon (left) and longitudinal virtual photon (right), plotted as a function of $p_T^2$ for $\xi=0.3$ and $Q^2=$ 0, 1, 5, and 10~GeV$^2$.
 }
\label{Telectro}
\end{figure}
%

%
\begin{figure}[htb]
\centerline{\epsfig{file=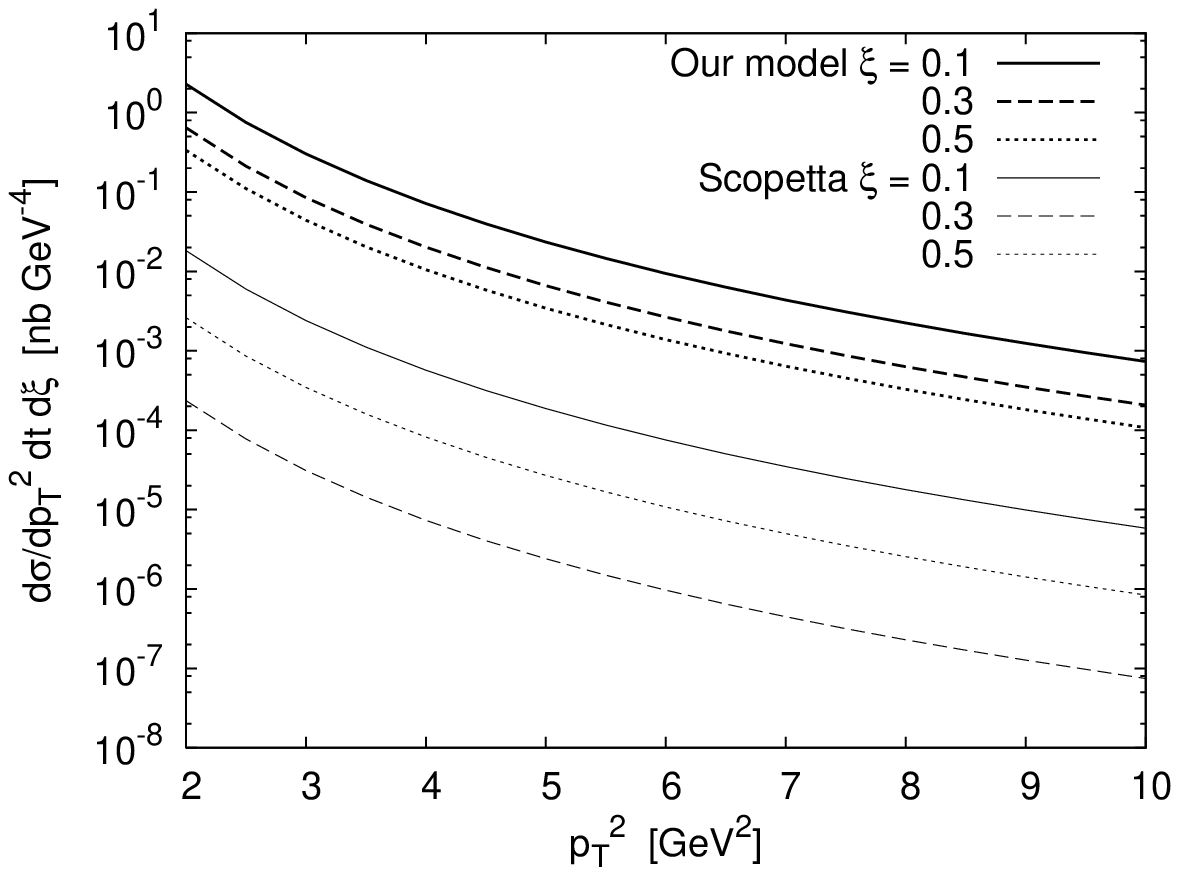,width=\wid}}
\caption{\small
Comparison of photoproduction of $\rho^0_L$,
$\rho^+_T$ computed using our model for $H_T$ and the model of \cite{Scop}.
}
\label{Tphoto_comparison}
\end{figure}
%

Let us now estimate the rates for the production of transversely 
polarized $\rho^+_T$ meson and 
discuss the feasibility of measuring the transversity GPD in the 
proposed process. 
We first note  that 
since the 
analytic expression (\ref{Hpole}) for the chiral-odd GPD is rather simple
we can compute, in the case of photoproduction, 
the integral over $u$ in (\ref{CON}) analytically. 
One obtains\footnote{We emphasize that this result holds for our simple model 
estimate of the chiral-odd GPD, and not for a more complete treatment.} 
\be
\label{oddexact}
{\cal M}^{\gamma \,p\,\to \rho_L^0\, \rho^+_T\,n}\left.\right|_{Q^2=0}=
 \sin \theta  \frac{ 216\,\pi^3 \, s \,\alpha_s^2 \, e\, C_F}{N_c^2}
 \frac{g_{b_1 N N} f_\rho^T f_\rho^0 f_{b_1}^T \langle k_\perp^2\rangle}{M_p m_{b_1}^2}
 \sqrt\frac{1-\xi}{1+\xi}
\frac{1}{|\p|^5},
\ee
and therefore for the cross section
\be
\label{csecodd}
\frac{d\sigma}{dp_T^2\,dt\,d\xi} = 
\frac{729\,\pi^4 \, \alpha_s^4 \, \alpha_{em}\, C_F^2}{N_c^4} 
 \frac{\left[g_{b_1 N N} f_\rho^T f_\rho^0 f_{b_1}^T \langle k_\perp^2\rangle\right]^2}{M_p^2 m_{b_1}^4}
\frac {\sin^2 \theta} {\xi(1+\xi) |\p|^{10}}.
\ee
For $Q^2 >0$ 
we evaluate the integrals 
numerically.
The cross section 
 for the transversely polarized 
$\rho$ case has a characteristic 
$\sin^2 \;\theta$ dependence (see Eq.~(\ref{CON})), which we do not explore here.

The differential cross sections for real and virtual  
photoproduction  of 
 transversely 
polarized $\rho^+$, based on  Eq.~(\ref{csecodd}), are shown in Figs.~\ref{Tphoto}, \ref{Tphotoxi} and \ref{Telectro}. 
As for the longitudinal $\rho^+_L$ case, we 
conclude from them that the photoproduction rate is much larger than the 
electroproduction rate (note again that we did not include the 
additional suppression factor coming from the virtual photon flux).
Fig.~\ref{Tphoto_comparison} shows a comparison of the corresponding 
photoproduction cross sections computed using our model (\ref{Hpole}) 
for $H_T$ and the bag model estimate of \cite{Scop}. 
Fig.~\ref{Tphoto_comparison} illustrates the strong sensitivity of
the production rate with respect to the used transversity GPD, and also the different $\xi$ dependence of the two.

\section{Discussion and conclusions}

Up to now we have shown that the process discussed here has a sizable
cross section with respect to the quite large luminosities of current
and projected high energy electron accelerators, In addition, one
should ask the question of the detection efficiency of existing or planned
experiments in the kinematical region where the two vector mesons are produced.
A crucial point in favor of a good detection
efficiency of our process is the fact that the two vector mesons
are produced with large transverse momentum. Let us detail this
statement.
Detecting a quasi-forward diffractively produced $\rho^0$ is now a
routine work for experimentalists at HERA and COMPASS, and we do not
expect any problem on this point in a forthcoming electron--proton
collider. The fact that we require that this $\rho^0$ has a finite
transverse momentum
of order 1 GeV or more will improve the detection efficiency.
Consequently, regarding the first (most rapid) vector meson,
the two charged pions emerging from it should be
well measured in modern detectors.

The new aspect of the process that
we have studied is the necessity to measure the charged $\rho$ meson
which is not diffractively produced and which travels in the
vicinity of the proton. Since one needs to measure its
polarization, a reasonable angular coverage of the two outgoing pions
is required. Without performing a detailed study of the acceptance
corrected cross section for a given experimental setup, we can
however make some definite remarks. The fact that the $\rho^0$ meson has a
sizable transverse momentum with respect to the quasi-real photon
(i.e.\ lepton) beam, and the fact that the transversity GPD is peaked
at small values of $t$ favors the case of a $\rho^+$ meson with a
similarly sizable transverse momentum (say greater than 1 GeV),
which is shared between both emerging pions. This is good news with
respect to the detection of these pions. To go further, one should
distinguish between fixed target experiment such as Compass and
collider experiments such as H1 or ZEUS. In the first case, the
longitudinal momentum of the $\rho^+$ meson is (in the laboratory
frame) of the order of $\sqrt{ s_{\gamma p} }$ which is a few GeV, so
that the emission angles of the $\rho^+$ and hence those of the two $\pi$
mesons are within the acceptance of the spectrometer. In
the collider kinematics, the efficiency may not be as good since the
vector meson is boosted along the direction of the proton beam. Note
however that its longitudinal momentum is proportional to $\xi$,
so that there is a small $\xi$ region where the emerging mesons should
not be hidden in the dead cone around the proton beam. These positive
remarks do not mean that a detailed analysis is not needed, neither
that special care is not required to improve as much as possible the
detection efficiency, but they allow us to be quite optimistic about
the feasibility of the proposed experiment.

We believe that the possible eRHIC machine may become the best place where the process discussed here could be measured, provided experimental setups allow a large angular coverage, ensuring a sufficient detection efficiency and a good control of exclusivity.  The JLab CLAS-12 upgrade probably will have good enough detection efficiency for observing the two rho mesons, but only for relatively low $p_T$ of the order of 1--1.5 GeV. Moreover the smaller energy available prevents the theoretical framework used here from being adequate and  one needs to supplement our studies by adding contributions coming from other polarization states of exchanged gluons and the ones coming from quark exchanges.
 This we leave for a future work.

The experimental measurement of the transversity GPD can thus supplement the intense present activity to unravel 
the transverse spin structure of the nucleon. 
An experimental determination of the transversity GPD $H_{T}$
seems feasible in photo- or electroproduction at high energies 
if the accelerator
luminosity is of the order of what is anticipated
at a future high-energy electron-proton collider.

\section*{Acknowledgments}
We thank S.\ Scopetta for providing us with the parametrization of the GPD from \cite{Scop}. We acknowledge useful discussions on experimental prospects with H.\ Avagyan, E.\ Burtin, J.\ Ciborowski, F.\ Kunne, A.\ Sandacz, L.\ Schoeffel and J.\ Ukleja. We also thank P.\ H\"agler, D.\ Ivanov and O.V.\ Teryaev for helpful discussion.
This work was supported in part by the Polish Grant 1 P03B 028 28,  
the French--Polish scientific agreement Polonium, the Joint Research
Activity ``Generalised Parton Distributions'' of the European I3 program
Hadronic Physics, contract RII3-CT-2004-506078, and the Swedish Research Council.
L.Sz.\ is a Visiting Fellow of the Fonds National pour la Recherche 
Scientifique (Belgium).



\begin{thebibliography}{99}

\bibitem{tra}
  J.~P.~Ralston and D.~E.~Soper,
  Nucl.\ Phys.\ B {\bf 152}, 109 (1979);
\\
  X.~Artru and M.~Mekhfi,
  Z.\ Phys.\ C {\bf 45}, 669 (1990);
\\
  J.~L.~Cortes, B.~Pire and J.~P.~Ralston,
  Z.\ Phys.\ C {\bf 55}, 409 (1992);
\\
  R.~L.~Jaffe and X.~D.~Ji,
  Phys.\ Rev.\ Lett.\  {\bf 67}, 552 (1991).
  
\bibitem{review}
  V.~Barone, A.~Drago and P.~G.~Ratcliffe,
  Phys.\ Rept.\  {\bf 359}, 1 (2002);\\
  M.~Anselmino,
  arXiv:hep-ph/0512140.

\bibitem{COGPD}
A.~V.~Belitsky and D.~M\"uller,
Phys.\ Lett.\ B {\bf 417}, 129 (1998);
\\
P.~Hoodbhoy and X.~D.~Ji,
Phys.\ Rev.\ D {\bf 58}, 054006 (1998);
\\
M.~Diehl,
Eur.\ Phys.\ J.\ C {\bf 19}, 485 (2001);
\\
M.~Diehl, T.~Gousset and B.~Pire,
Phys.\ Rev.\ D {\bf 59}, 034023 (1999);
\\
J.~C.~Collins and M.~Diehl,
Phys.\ Rev.\ D {\bf 61}, 114015 (2000);
\\
  M.~Diehl and P.~H\"agler,
  Eur.\ Phys.\ J.\ C {\bf 44}, 87 (2005).



\bibitem{IPST}
  D.~Y.~Ivanov, B.~Pire, L.~Szymanowski and O.~V.~Teryaev,
  Phys.\ Lett.\ B {\bf 550}, 65 (2002)



\bibitem{BFKL} 
L.~N.~Lipatov,
Sov.\ J.\ Nucl.\ Phys.\  {\bf 23}, 338 (1976);
\\
E.~A.~Kuraev, L.~N.~Lipatov and V.~S.~Fadin,
Sov.\ Phys.\ JETP {\bf 44}, 443 (1976); 
\\
E.~A.~Kuraev, L.~N.~Lipatov and V.~S.~Fadin,
Sov.\ Phys.\ JETP 
{\bf 45}, 199 (1977);
\\
I.~I.~Balitsky and L.~N.~Lipatov,
Sov.\ J.\ Nucl.\ Phys.\ {\bf 28}, 822 (1978).



\bibitem{BalBr}
  P.~Ball and V.~M.~Braun,
  Phys.\ Rev.\ D {\bf 54}, 2182 (1996).



\bibitem{Diehl:2001pm}
  M.~Diehl,
  Eur.\ Phys.\ J.\ C {\bf 19}, 485 (2001).


\bibitem{Radyushkin}
  A.~V.~Radyushkin,
  Phys.\ Rev.\ D {\bf 59}, 014030 (1999).


\bibitem{Dterm}
  M.~V.~Polyakov and C.~Weiss,
  Phys.\ Rev.\ D {\bf 60}, 114017 (1999).


\bibitem{GPV}
  K.~Goeke, M.~V.~Polyakov and M.~Vanderhaeghen,
  Prog.\ Part.\ Nucl.\ Phys.\  {\bf 47}, 401 (2001).


  
\bibitem{MRST}
  A.~D.~Martin, R.~G.~Roberts, W.~J.~Stirling and R.~S.~Thorne,
  Phys.\ Lett.\ B {\bf 604}, 61 (2004).

\bibitem{JJ}
  R.~L.~Jaffe and X.~D.~Ji,
  Phys.\ Rev.\ D {\bf 43}, 724 (1991).


\bibitem{Scop}
  S.~Scopetta,
  Phys.\ Rev.\ D {\bf 72}, 117502 (2005).
See also  B.~Pasquini, M.~Pincetti and S.~Boffi,
Phys.\ Rev.\ D {\bf 72}, 094029 (2005).


\bibitem{Lattice}
  M.~G\"ockeler {\it et al.}  [QCDSF Collaboration],
  Phys.\ Lett.\ B {\bf 627}, 113 (2005).

  


\bibitem{MPR}
  L.~Mankiewicz, G.~Piller and A.~Radyushkin,
  Eur.\ Phys.\ J.\ C {\bf 10}, 307 (1999).

\bibitem{GG}
  L.~P.~Gamberg and G.~R.~Goldstein,
  Phys.\ Rev.\ Lett.\  {\bf 87}, 242001 (2001).


\end{thebibliography}
\end{document}